# Excess wing in glass-forming glycerol and LiCl-glycerol mixtures detected by neutron scattering

S. Gupta,[1] N. Arend,[1,2] P. Lunkenheimer,[3] A. Loidl,[3] L. Stingaciu,[1] N. Jalarvo,[1] E. Mamontov,[4] and M. Ohl[1]

[1]*Jülich Centre for Neutron Science JCNS, Forschungszentrum Jülich GmbH, Outstation at SNS, 1 Bethel Valley Road, Oak Ridge, TN 37831, USA*

[2]*Jülich Centre for Neutron Science JCNS, Forschungszentrum Jülich GmbH, Outstation at MLZ, Lichtenbergstraße 1, 85747 Garching, Germany*

[3]*Experimental Physics V, Center for Electronic Correlations and Magnetism, University of Augsburg, 86135 Augsburg, Germany*

[4]*SNS, Neutron Sciences Directorate, Oak Ridge National Laboratory (ORNL), 1 Bethel Valley Road, Oak Ridge, TN-37831, USA*

The relaxational dynamics in glass-forming glycerol and glycerol mixed with LiCl is investigated using different neutron scattering techniques. The performed neutron spin-echo experiments, which extend up to relatively long relaxation-time scales of the order of 10 ns, should allow for the detection of contributions from the so-called excess wing. This phenomenon, whose microscopic origin is controversially discussed, arises in a variety of glass formers and, until now, was almost exclusively investigated by dielectric spectroscopy and light scattering. Here we show that the relaxational process causing the excess wing also can be detected by neutron scattering, which directly couples to density fluctuations.

## I. INTRODUCTION

The glass transition is one of the most fascinating mysteries in condensed matter physics. To understand the non-canonical increase of viscosity when a liquid becomes a glass, the investigation of molecular dynamics by scattering methods or dielectric spectroscopy is a common approach. Broadband susceptibility spectra obtained in this way reveal a complex variety of dynamic processes [1,2,3,4,5]. This includes the so-called $\alpha$ relaxation, which is the structural relaxation process governing the viscosity. In the imaginary part of the susceptibility $\chi''$ (which corresponds to the loss $\varepsilon''$ in the dielectric case) usually it shows up as a prominent peak located at a frequency $\nu_\alpha$, which is wider than predicted by the Debye theory due to the typical dynamic heterogeneity of supercooled liquids [1,6]. However, there are also a number of additional, faster dynamic processes, whose microscopic origin is still controversially discussed. Understanding these processes seems to be a prerequisite to achieve a better understanding of the glass transition and the glassy state of matter in general.

Consequently, in recent years a lot of experimental work has been devoted to these phenomena. The most prominent of these faster processes are the excess wing [2,3,4,5,7,8], the Johari-Goldstein (JG) $\beta$ relaxation [3,5,9,10,11], and the fast $\beta$ relaxation [2,3,4,5,8,12,13,14,15,16]. Information on the latter, typically showing up at frequencies between several 100 MHz and 1 THz, was mainly collected by neutron and light scattering [13,14,15,16] and also by dielectric spectroscopy [2,3,8,11,17]. In contrast, the excess wing and JG relaxation where almost exclusively investigated by dielectric spectroscopy and, partly, also by light scattering [2,3,4,5,7,8,9,10,11]. In $\chi''$ spectra, these phenomena are usually found at relatively low frequencies ($\nu \lesssim 10^8$ Hz) only, inaccessible by most neutron scattering experiments. In the spectra, they show up as an excess contribution to the high-frequency flank of the $\alpha$ peak (thus termed "excess wing" [8,18]) or as a second relaxation peak. An increase of temperature in principle should lead to a shift of these processes to higher frequencies. However, as the $\alpha$-relaxation time $\tau_\alpha \approx 1/(2\pi\nu_\alpha)$ generally has much stronger temperature dependence, these spectral features tend to merge with the $\alpha$-relaxation peak when the temperature significantly exceeds the glass temperature $T_g$. Therefore, they cannot be detected at the relatively high frequencies covered by most neutron scattering experiments.

Generally, the microscopic origin of excess wing and JG relaxation is controversially discussed. It even is unclear if they are caused by the same or different processes. While it was demonstrated that the excess wing arises from a secondary relaxation peak, partly submerged under the dominating $\alpha$ peak [19,20,21,22], it is not so clear if this peak is due to the JG relaxation [19,20,23] or another, separate relaxation process [10,24,25]. The dielectric measurements of dipolar molecular glass formers that are usually performed to investigate the excess wing are mainly sensitive to orientational fluctuations. The latter also play a strong role in light-scattering experiments [26,27]. Thus, detecting the excess wing with neutron scattering, which directly couples to density fluctuations, would be an important piece of information in the search for the microscopic origin of this phenomenon.

In the present work, results from neutron spin echo (NSE) spectroscopy are presented, extending up to relatively long relaxation time scales of the order of 10 ns, which corresponds to a lower frequency limit of about $10^7$ Hz. Recently, indications for an excess wing in an aqueous solution of LiCl where found using this method [28]. Applying NSE spectroscopy to glass-forming glycerol and glycerol mixed with LiCl, we look for contributions of the excess wing at low temperatures. Excess wing and $\alpha$ relaxation were shown to become more separated when LiCl is added to glycerol, which makes this system well suited to study this process [17,29]. In addition, the $\alpha$ relaxation is investigated in detail, which is complemented by results from neutron backscattering (BS) experiments, covering a relaxation-time range of the order of $10^{-11}$ - $10^{-9}$ s.

## II. EXPERIMENTAL DETAILS

For the BS experiments, glycerol of ≥99.5% purity from Sigma-Aldrich was used. For the NSE experiments, deuterated glycerol with a purity of 98% and 99% deuteration level was purchased from Cambridge Isotope Laboratories. For the preparation of the LiCl solution with a molar concentration of 4%, anhydrous, ultra-dry lithium chloride powder (purity 99.95%) from Alfa-Aesar was used.

The quasi-elastic neutron scattering experiments were performed at the Spallation Neutron Source (SNS), Oak Ridge National Laboratory (ORNL) using the BS spectrometer BASIS [30] and the Spin Echo spectrometer SNS-NSE [31]. The measurement configuration at BASIS was the standard one, with 3.4 µeV energy resolution (full-width at half maximum, $Q$-averaged value) and the dynamic range of ±100 µeV selected for the data analysis. At the NSE experiment an incoming wavelength band from 5 to 8 Å$^{-1}$ was used with 42 time channels for the time-of-flight data acquisition. This allowed to access a dynamic range of 2 ps $\leq t \leq$ 10 ns at a given momentum transfer. We measured the dynamics of deuterated glycerol molecules with and without LiCl. Corrections were performed using resolution data from a TiZr sample and background from the empty cell. We used aluminum sample containers sealed with indium wires. The data reduction was performed with the standard ECHODET software package of the SNS-NSE instrument. A short description of the NSE spectroscopy is given in the Supplemental Material [32]. For further details the reader is referred to Refs. [33,34,35].

## III. RESULTS AND DISCUSSION

Figure 1 shows dielectric loss spectra from Ref. [17], measured for the same glass formers as investigated in the present work, namely pure glycerol and glycerol mixed with 4% LiCl. The data, shown for three selected temperatures, cover a frequency range of 12 decades. For pure glycerol, the spectra are dominated by the $\alpha$-relaxation peak, whose shift towards lower frequencies with decreasing temperature mirrors the glass transition [2,3]. Adding LiCl leads to the emergence of a $1/\nu$ contribution at low frequencies, partly superimposing the $\alpha$ peaks. It arises from ionic charge transport as discussed in detail in Ref. [36]. In addition, the peak position shifts to lower frequencies, i.e. the relaxation time $\tau_\alpha$ increases with increas-ing ion concentration. This was ascribed to interactions between the glycerol molecules and ions, causing a reduced molecular mobility [36]. At low temperatures, in glycerol an excess wing is known to appear [2,3,7]. For example, at 213 K in Fig. 1 it is seen between about $10^6$ and $10^9$ Hz for pure glycerol. In the region between about $10^9$ and $10^{12}$ Hz a shallow minimum shows up. It has been interpreted in terms of the fast $\beta$ process [2,3,17] as predicted by the mode-coupling theory [12]. There are also alternative explanations of the minimum, e.g., by a constant-loss contribution expected within the extended coupling model [37]. Finally, at around 1 THz, a broad peak shows up, which corresponds to the boson peak [2,3,5,38]. The resonance-like features beyond $10^{13}$ Hz can be ascribed to intramolecular excitations. In Fig. 1, the frequency/time regions that in principle can be accessed by the NSE and BS setups employed in the present work are indicated. Obviously, the range of the NSE experiment extends to sufficiently low frequencies to allow for the detection of the excess wing, at least for the lowest temperature of 213 K.

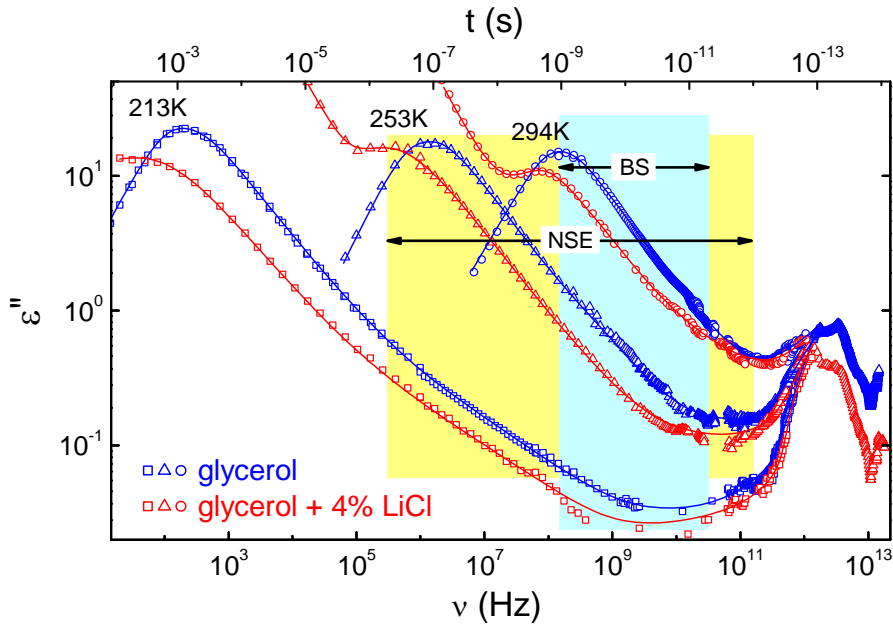

FIG. 1. (Color) Comparison of broadband dielectric loss spectra of glycerol and glycerol with 4% LiCl salt, shown for three typical temperatures as published in [17]. The lines are guides to the eye. For readability, in the boson-peak region ($\nu >$ 1 THz) the results are shown for 253 K, only. The frequency and time scales accessible by the two neutron scattering techniques of the present work (NSE and BS) are indicated in the figure (lower and upper scales, respectively).

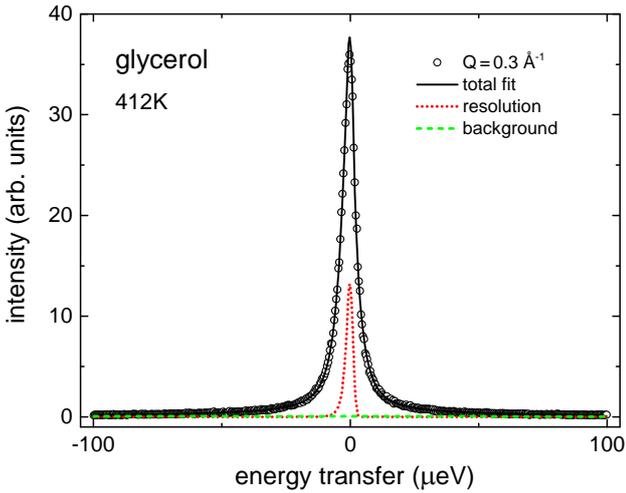

FIG. 2. (Color online) BS data (symbols) for pure glycerol at 412K, measured at $Q = 0.3$ Å$^{-1}$. The solid line is a fit with Eqs. (1) and (2). The contributions from the resolution function and linear background term are indicated by the dotted and dashed lines, respectively.

As a typical example for the BS data, Fig. 2 shows the results obtained for pure glycerol, measured at 412 K and $Q = 0.3$ Å$^{-1}$. All obtained BS results were fitted for each $Q$ value using the following expression [39]:

$$I(Q,E) = \{X(Q)\delta(E) + [1 - X(Q)] S(Q,E)\} \otimes R(Q,E) + B(Q,E) \quad (1)$$

Here $\delta(E)$ is a delta function centered at zero energy transfer, $X(Q)$ represents the fraction of the elastic scattering, $B(Q,E)$ is the linear background term, $S(Q,E)$ is the dynamic struc-ture factor in frequency space, and $R(Q,E)$ is the resolution function as measured for the sample at low temperature (20 K). The elastic term represents any elastic background arising from the sample holders, sample environment, and the spectrometer background as described by Mamontov *et al.* [39]. We use the following expression for the one-Lorentzian dynamic structure factor [40]:

$$S(Q,E) = \frac{1}{\pi} \frac{\Gamma(Q)}{E^2 + \Gamma^2(Q)} \quad (2)$$

The parameter $\Gamma(Q)$ represents the half-width at half maximum of the BS signal. As revealed by Fig. 2, the fitting function (solid line) describes the data quite accurately, which also is the case for the results at other $Q$ values. The dotted line indicates the contribution from the resolution function. The linear background term, shown by the dashed line, is of the order of 0.05 and thus appears close to zero in the scale of Fig. 2.

The half width $\Gamma$ in Eq. (2) can be related to the relaxation time $\tau$, associated with the long-range translational diffusion coefficient $D$, as $\tau = \hbar/\Gamma(Q)$ [40]. From this diffusive behavior, we determined the structural relaxation time $\tau_0$ assuming $\tau = \tau_0 Q^{-x}$, where $x = 2 \pm 0.3$ in our case, indicating diffusive relaxation. A similar analysis was performed for all concentrations and temperatures. The extracted values of $\tau_0$ vary between 0.02 and 2.2 ns. For low temperatures, $T < 270$ K, $\tau$ lies beyond the instrumental resolution and hence $\tau_0$ could not be determined. The obtained results for $\tau_0$ at different LiCl concentrations will be compared with the dielectric and NSE results in the further course of this work.

NSE spectroscopy measures the normalized dynamic structure factor $S(Q,t)/S(Q,0)$ as a function of Fourier time $t$ at a given momentum transfer $Q$. Figure 3 shows $S(Q,t)/S(Q,0)$ for two samples, pure glycerol and 4% LiCl in glycerol, for different temperatures at $Q = 0.074$ Å$^{-1}$. Using the SNS-NSE instrument, we performed coherent neutron scattering with a lower contribution of incoherent scattering (see Eq. (A4) in Supplemental Material [32]). Thereby, we choose $Q = 0.074$ Å$^{-1}$ to observe the translational collective dynamics of the glycerol molecules by accessing the density correlation function. Our goal was to compare the relaxation times obtained in the NSE experiment with those of the structural relaxation and the underlying excess wing of the systems as obtained from dielectric spectroscopy. Figure 3 reveals a decay of the dynamic structure factor from an upper to a lower plateau value, which shifts to longer times when the temperature is lowered. This is the typical signature of relaxational behavior in time-domain data. The values of the plateaus found at short times are smaller than 1. This clearly illustrates the presence of initial faster dynamics (< 1 ps) arising, e.g., from the fast $\beta$ process and microscopic dynamics. At first glance, it seems likely that the structural $\alpha$-relaxation dominates the detected decay. Indeed, NSE data on pure glycerol, measured at $T \geq 270$ K and covering a somewhat smaller time range, could be reasonably interpreted in this way [41].

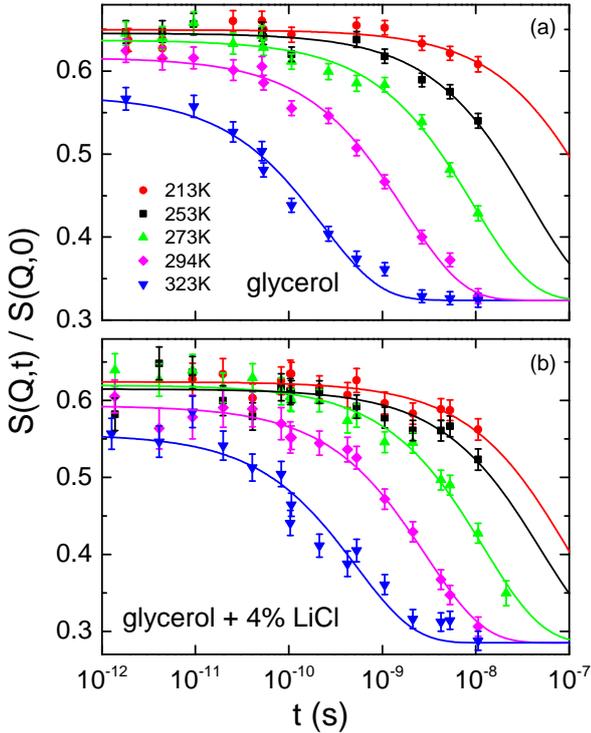

FIG. 3. (Color online) Normalized dynamic structure factor $S(Q,t)/S(Q,0)$ at $Q = 0.074$ Å$^{-1}$ for (a) pure glycerol and (b) glycerol with 4% LiCl, measured at different temperatures as indicated in the legend. The solid lines are fits using Eq. (3) with an additional offset parameter as described in the text.

To fit time-domain data of the $\alpha$ relaxation, the stretched-exponential function, also termed Kohlrausch-Williams-Watts (KWW) function [42], is commonly employed. It is defined by:

$$\Phi(t) = A\exp\left[-\left(\frac{t}{\tau_{\text{KWW}}}\right)^{\beta_{\text{KWW}}}\right] \quad (3)$$

In case of neutron scattering, $\Phi$ corresponds to the normalized dynamic structure factor $S(Q,t)/S(Q,0)$ and $A$ is the Debye-Waller factor. The exponent $\beta_{\text{KWW}}$ determines the stretching of the exponential function; for $\beta_{\text{KWW}} = 1$, exponential decay is recovered. For $\beta_{\text{KWW}} < 1$, the decay of $\Phi(t)$ becomes smeared out, which can be ascribed to a distribution of relaxation times [6]. It should be noted that the Fourier transform of the KWW function leads to a qualitatively similar spectral shape as the Cole-Davidson function, which was shown to provide reasonable fits of the $\alpha$ relaxation in dielectric spectra of these glass formers [3,36]. To enable a comparison of relaxation times obtained from different fit functions, an average relaxation time can be calculated for each relaxation curve. For the KWW case, it is defined by $\langle\tau\rangle = \tau_{\text{KWW}}/\beta_{\text{KWW}}\,\Gamma(1/\beta_{\text{KWW}})$, where $\Gamma$ denotes the Gamma function [43]. For the CD function, it is given by $\langle\tau\rangle = \tau_{\text{CD}}\beta_{\text{CD}}$ [2].

The lines in Fig. 3 are fits using the KWW function, Eq. (3), with an additional offset parameter. The latter accounts for the fact that the dynamic structure factor does not approach zero at long times (clearly revealed at the highest temperatures shown in Fig. 3), which can be ascribed to elastic scattering from the background including the sample environment and instrument background. For both glass formers, the long-time limit of $S(Q,t)/S(Q,0)$ was determined from the data at the highest investigated temperature. This value was also used for the fits of the data at lower temperatures, where the long-time plateau is not well defined. The stretching parameter $\beta$ was deduced with a similar procedure as described in Ref. [41], leading to $\beta = 0.7$ for all concentrations. This agrees with the value found for pure glycerol reported by Wuttke et al. [41]. From the fits, the Debye-Waller factor $A$ in Eq. (3) was found to decrease exponentially with temperature as expected for a harmonic solid.

The obtained results for the temperature-dependent average $\alpha$-relaxation times are shown in Fig. 4, plotted in an Arrhenius representation. Here the closed circles correspond to the results from NSE spectroscopy at $T > 213$ K and the open circles are literature data from dielectric spectroscopy [29,36] (the NSE data at 213 K represent a special case and will be treated below). For most temperatures only a part of the decay curves could be probed with the available experimental setup (Fig. 3), which leads to uncertainties in $\langle\tau\rangle$. In light of this fact, the agreement of dielectric and neutron scattering data is reasonable. For pure glycerol, it is well known [15,44,45] that the $\alpha$-relaxation times determined by different experimental methods, e.g., dielectric spectroscopy, light, and neutron scattering, are quite similar. Figure 4 also includes the relaxation times from the BS measurements. While they are of comparable order of magnitude, partly there are deviations from $\tau_\alpha$ deduced by NSE and dielectric spectroscopy. At the highest temperatures this is a consequence of the system dynamics being somewhat too fast for the accessible dynamic range of the BS spectrometer. The opposite effect is evident at the lowest temperature of the experiment, where the BS spectrometer yields somewhat shorter relaxation times compared to the NSE and dielectric results, due to the dynamics slowing down beyond the spectrometer resolution. Moreover, it cannot be excluded here that contributions from processes faster than the $\alpha$ relaxation (e.g., the fast $\beta$ relaxation) lead to a broadening of the observed $S(Q,E)$ peaks, and thus a reduction of the deduced relaxation time. The error bars shown in Fig. 4 account for these effects.

As mentioned above, the data at 213 K shown in Fig. 3 represent a special case and cannot be interpreted in the same manner as those measured at higher temperatures. This becomes obvious from the fact that the relaxation times deduced from the fits of these data (closed triangles in Fig. 4) strongly deviate from the dielectric $\alpha$-relaxation times (open circles). At this temperature, the average $\alpha$-relaxation times from dielectric spectroscopy are $8\times10^{-4}$ s for pure glycerol and $3\times10^{-3}$ s for 4% LiCl [29,36] (cf. Fig. 4). This is several orders of magnitude larger than the maximum times covered by the NSE experiments ($10^{-8}$ s). Therefore, the decay of the dynamic structure factor observed at 213 K (Fig. 3) clearly cannot stem from the $\alpha$ relaxation (see dashed line in Fig. 5 for an illustration of this fact). This also becomes obvious if comparing the $\alpha$-peak positions at 213 K in Fig. 1 with the time range covered by the NSE experiment. Figure 1 also reveals that at 213 K the excess wing is the strongest contribution in the time range of about $10^{-9} - 10^{-8}$ s, where this decay is observed in Fig. 3. All these findings provide strong indications that at 213 K, instead of the $\alpha$ relaxation, the excess-wing relaxation is detected by the performed neutron scattering experiments. Notably, for both glass formers the relaxation times from NSE at 253 K tend to be somewhat smaller than the dielectric $\alpha$-relaxation times. Taking into account the error bars of the NSE data, both data sets still seem to be compatible. However, one may also speculate that already at 253 K the excess wing starts to play some role and leads to a slightly faster decay of the time-dependent dynamic structure factor.

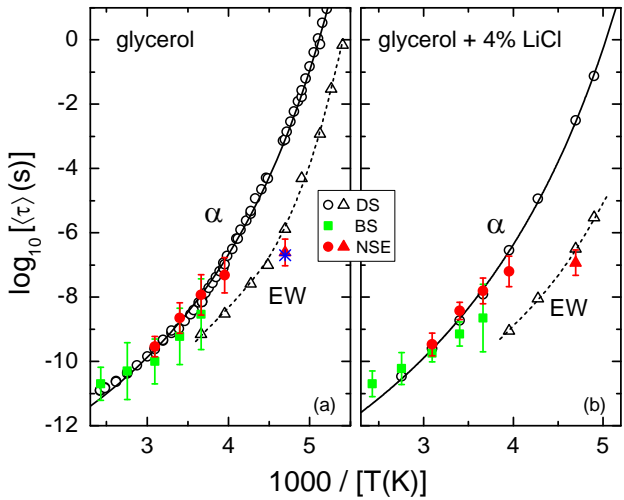

FIG. 4. (Color online) Relaxation time maps in Arrhenius representation for the investigated glass formers. The open circles show the average $\alpha$-relaxation times and the open triangles those of the excess-wing relaxation as determined from dielectric spectroscopy [2,3,23,29,36]. The solid lines are fits with the Vogel-Fulcher-Tammann function [29,36]; the dashed lines are guides to the eye. The closed squares show the relaxation times determined from the BS experiments. The closed circles and triangles represent the relaxation times determined from the NSE experiments shown in Fig. 3. The values at 213 K (closed triangles) match the excess-wing relaxation times deduced by dielectric spectroscopy. The star in (a) shows $\tau_{EW}$ used for the calculation of the double-step function in Fig. 5.

In Fig. 4, in addition to the $\alpha$-relaxation times, also the characteristic times of the excess-wing relaxation $\tau_{EW}$ deduced from dielectric spectroscopy are shown (open triangles) [23,29]. One should have in mind that these data have a relatively high uncertainty because no peak but only a second power law is seen in the spectra [23,29]. Nevertheless, for both materials the closed triangles in Fig. 4, showing the relaxation times deduced from the NSE data at 213 K, are of similar magnitude as $\tau_{EW}$ from dielectric spectroscopy (open triangles). This nicely corroborates the notion that at 213 K indeed the excess wing was detected by the performed neutron scattering experiments. One should note that, within this scenario, in principle the NSE data at 213 K should be fitted by a two-step relaxation function accounting for excess wing and $\alpha$ relaxation. However, due to the strong separation of relaxation times, a fit with a single KWW function as performed for 213 K can be assumed to provide at least a rough estimate of $\tau_{EW}$.

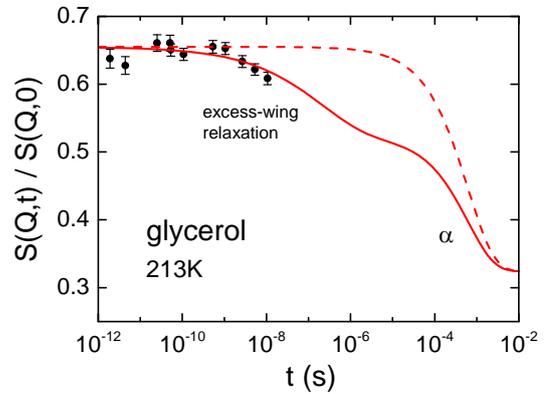

FIG. 5. (Color online) Normalized dynamic structure factor at $Q = 0.074$ Å$^{-1}$ and 213 K for pure glycerol as already shown in Fig. 3(a) (circles). The solid line was calculated using the combination of two KWW functions, Eq. (4), as described in the text. The dashed line is a single KWW function with the same average relaxation time as deduced from dielectric data [3].

The application of a two-step relaxation function is illustrated in Fig. 5 showing the 213 K data for pure glycerol. The solid line was calculated using the Williams product ansatz combining two KWW functions [46], namely:

$$\Phi(t) = A\left\{p + (1-p)\exp\left[-\left(\frac{t}{\tau_{EW}}\right)^{\beta_{EW}}\right]\right\}\exp\left[-\left(\frac{t}{\tau_\alpha}\right)^{\beta_\alpha}\right] \quad (4)$$

Here $p$ is the relative strength factor of the $\alpha$ process ($p = 1$ leads to a single step arising from the $\alpha$ relaxation only). For $\tau_\alpha$, we used $6.2\times 10^{-4}$ s, which corresponds to the same average relaxation time as deduced from dielectric spectroscopy [3]. The width parameter of the $\alpha$ relaxation was set to $\beta_\alpha = 0.7$, the same value as used for the fits in Fig. 3(a). Again, an additional offset parameter was introduced, which was chosen to be of the same magnitude as for the high-temperature data. The experimental data in Fig. 5 only show the onset of the decay of the dynamic structure factor. There-fore, it is clear that, even if fixing some of the parameters as noted above, a meaningful fit of the data with the two-step function of Eq. (4) cannot be performed. Especially the factor $p$ cannot be unequivocally determined. Nevertheless, the solid line in Fig. 5, calculated for $p = 0.6$, at least demonstrates that the experimental data are fully consistent with such a two-step scenario involving an excess-wing and $\alpha$ relaxation. Moreover, the dashed line in this plot, calculated for $p = 1$ and using the $\alpha$-relaxation time from dielectric spectroscopy [3], demonstrates that the experimental data are clearly inconsistent with pure $\alpha$ response. Glycerol with 4% LiCl reveals similar behavior: Notably, for this glass former the $\alpha$ relaxation at 213 K takes place at even longer times [29,36] (cf. Fig. 4) and the observed decay at $10^{-9}$ - $10^{-8}$ s [Fig. 3(b)] clearly cannot be understood without assuming a contribution from the excess-wing relaxation.

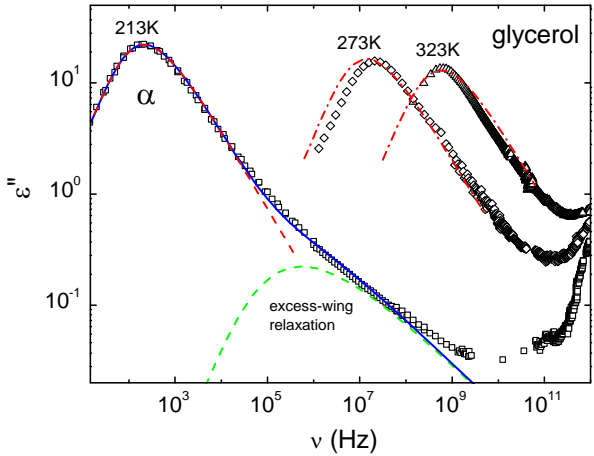

FIG. 6. (Color online) Broadband dielectric loss data of pure glycerol for three selected temperatures [3]. The dashed lines show the scaled Fourier transforms of the two KWW functions giving rise to the two-step decay shown by the solid line in Fig. 5, which was calculated by Eq. (4). The solid line shows the sum of these two peaks. The dash-dotted lines are the Fourier transforms of the KWW functions used to fit the NSE time-domain data at 273 and 323 K in Fig. 3(a).

As indicated by the star in Fig. 4(a), the excess-wing relaxation time used for the calculation of the solid line in Fig. 5 is in reasonable accord with the results from dielectric spectroscopy. Further comparison with the dielectric data is provided in Fig. 6, showing the dielectric loss spectra for pure glycerol at three temperatures, including 213 K [3]. The dashed lines represent the Fourier transforms of the two KWW functions used for the calculation of the two-step decay shown in Fig. 5, both leading to well-defined loss peaks. To compare them with the dielectric data, they were vertically shifted by applying a proper scaling factor. This is justified because the NSE data are normalized to unity and because the relative amplitude $p$ cannot be unequivocally deduced from the experimental data, as mentioned above. Moreover, the relative amplitudes of $\alpha$ process and excess wing can vary for different experimental probes [5]. The solid line in Fig. 6 shows the sum of the two peaks indicated by the dashed lines. Obviously, the two KWW functions that were proposed to explain the decay of the dynamic structure factor at 213 K in time domain (Fig. 5) are well consistent with the dielectric data measured in the frequency domain. The deviations observed at frequencies beyond about $10^8$ Hz arise from the fast $\beta$ process [2,3,16] or from a constant-loss contribution [37], which are out of the scope of the present work. One should bear in mind that here, for practical reasons, a KWW function was used to describe the excess-wing relaxation, leading to an asymmetric loss peak. In contrast, usually the symmetric Cole-Cole function is used for fitting secondary relaxations [10,11,20] but its Fourier transform does not lead to an analytical expression in time domain. However, this does not represent a major problem because the low-frequency wing of the excess-wing relaxation is superimposed by the much stronger $\alpha$ peak (see Fig. 6) and only its high-frequency wing significantly contributes to the overall curve.

As an example for the behavior at higher temperatures, the dash-dotted lines in Fig. 6 show the Fourier transforms of the KWW functions that were used to fit the neutron scattering data at 273 and 323 K [Fig. 3(a)], which are completely dominated by the $\alpha$ relaxation. The obtained peaks are consistent with the dielectric spectra, especially if considering the large uncertainty in the determination of the relaxation times as mentioned above.

## IV. SUMMARY AND CONCLUSIONS

In the present work, results from NSE spectroscopy on pure glycerol and glycerol mixed with LiCl, complemented by BS measurements, are provided. The NSE experiments enabled the investigation of the relaxational dynamics up to relatively long time scales of the order of 10 ns. In addition to the detailed investigation of the $\alpha$ relaxation, this allowed to search for indications of the excess wing, which plays a prominent role in the investigation of glassy dynamics by dielectric spectroscopy and light scattering. Indeed, at low temperatures we found clear indications for a decay of the dynamic structure factor than cannot be explained by the structural $\alpha$ relaxation. Instead, the experimental findings strongly point to contributions from a faster process. By comparison with broadband dielectric data on the same sample materials, we identify this fast process with the relaxation causing the excess wing. Thus, the excess-wing relaxation, mainly known from dielectric investigations of the reorientational dynamics in dipolar molecular glass formers, also is detectable by coherent neutron scattering, directly coupling to density-density fluctuations. As the excess wing and secondary relaxation process are universal features of glassy matter, this finding related to their microscopic origin also is of relevance to enhance our understanding of the peculiarities of the glass transition in general. While further experimental advances allowing investigations at even longer time scales are desirable, our results may open a new field of research on the excess wing and secondary relaxation processes using neutron scattering methods.

ACKNOWLEDGMENTS

We thank Prof. D. Richter, M. Monkenbusch, and R. Zorn for helpful discussions. Research conducted at ORNL's High Flux Isotope Reactor (and/or Spallation Neutron Source, as appropriate) was sponsored by the Scientific User Facilities Division, Office of Basic Energy Sciences, US Department of Energy. The work at the University of Augsburg was partly supported by the Deutsche Forschungsgemeinschaft via Research Unit FOR 1394.

———————————